\documentclass[12pt]{iopart}

\usepackage{amssymb}
\usepackage{graphicx}
\usepackage{xcolor}
\usepackage{color} 
\usepackage{url}
\begin{document}

\title[Simulated proton depth-dose curves and WER in cortical bone]{Simulation of depth-dose curves and water equivalent ratios of energetic proton beams in cortical bone}

\author{Ana María Zamora-Vinaroz$^1$, Pablo de Vera$^1$, \\ Isabel Abril$^2$, Rafael Garcia-Molina$^1$}

\address{$^1$Departamento de Física -- Centro de Investigación en Óptica y Nanofísica, Universidad de Murcia, 30100 Murcia, Spain}
\address{$^2$Departament de Física Aplicada, Universitat d’Alacant, 03080 Alacant, Spain}
\ead{pablo.vera@um.es}
\vspace{10pt}
\begin{indented}
\item[]\today
\end{indented}

\begin{abstract}
We have determined the depth-dose curve, the penetration range, and the water equivalent ratio (WER), for proton beams of clinical energies in cortical bone, by means of a detailed and accurate simulation that combines molecular dynamics and Monte Carlo techniques. The fundamental input quantities (stopping power and energy loss straggling) for the simulation were obtained from 
a reliable electronic excitation spectrum of the condensed-phase target, which takes into account the organic and mineral phases that form it. Our simulations with these inputs, that are in excellent agreement with the scarce data available for a cortical bone target, deviate from simulations performed using other stopping quantities, such as those provided by the International Commission on Radiation Units and Measurements (ICRU) in its widely used Report 49.  The results of this work emphasize the importance of an accurate determination of the stopping quantities of cortical bone in order to advance towards the millimetric precision for the proton penetration ranges and deposited dose needed in radiotherapy.
\end{abstract}

%
%
%
%
%

\section{Introduction} 

Nowadays, hadrontherapy  and, particularly, protontherapy, are being promoted because of their specific advantages in comparison to conventional radiotherapy. Hadrontherapy can be used to treat deep-seated tumours without damaging the surrounding healthy tissues or the surface tissues as happens with photon treatments. This is due to an increase of the radiobiological effectiveness, as well as the fact that heavy charged particles deposit their energy in the so-called Bragg peak, a narrow region at the end of their trajectories, where a pronounced dose peak is produced \cite{Kraft1990, Elsasser2009, Schardt2010, Limandri2014a, deVera2018}.

Treatment planning in hadrontherapy includes the simulation of therapeutic ion beams impinging on human tissues with energies of the order of tens and hundreds of MeV/u and a dose assessment. The latter is performed, for simplicity, by measuring the absorbed dose and the beam range in a patient phantom made of water, which is the main constituent of the human body. Finally, the water equivalent ratio (WER), as a nice tissue equivalence estimator \cite{deVera2014b, Bagheri2019}, allows to compute the beam range in biological tissues \cite{Burin2023}. 

As the Bragg peak is narrow, the possible uncertainties in range, which will exist in WER as well, could affect the treatment effectiveness and lead to a damage of the surrounding healthy tissues. Therefore, it is necessary to know accurately the energy deposition by swift protons in the tissues that will be sampled. This means that the electronic excitation spectrum of the target must be properly described taking into account the condensed-phase nature of the material. 

For the purpose of the present study, the biological materials are roughly classified between \textit{soft} and \textit{hard} (or skeletal) tissues \cite{ICRU46,Limandri2014a}. 
On the one hand, the former are usually assumed to be made of  liquid water, as it has been referred to be an excellent tissue-like phantom material for determination of absorbed dose \cite{ICRU1998}. This material has been studied in more detail, and even though the amount of experimental data on its stopping power for protons is rather limited \cite{Shimizu2009,Shimizu2010,Siiskonen2011},  there are several theoretical models capable of reproducing their main trends \cite{Dingfelder2000,Akkerman2001,Date2006,Emfietzoglou2009,GarciaMolina2009}. For other organic materials making up soft tissues, the stopping power (as well as other relevant energy loss quantities) can be calculated by a parameterisation of their excitation spectrum \cite{Tan2004}.

On the other hand, there is less information available for the stopping power of heavy charged particles in hard tissues, specially in bone. As there are very few experimental measurements in hard tissues \cite{Koehler1965}, the main way to compute its stopping power is the additivity of atomic stopping powers using Bragg's rule \cite{ICRU49,Berger2005}. However, 42\% of bone is collagen protein and the other 58\% is calcium hydroxyapatite (HAp), whose stochastic formula is Ca$_{10}$(PO$_4$)$_6$(OH)$_2$. The stopping power of HAp has been measured experimentally and compared with simulations based on the dielectric formalism \cite{Limandri2014a}, finding a nice agreement between theory and experimental data. Therefore, according to Bragg's rule, the stopping power of cortical bone can be computed by the weighted sum of the stopping powers of its mineral (HAp) and organic components, taking into account its condensed-phase nature.

The aim of the present work is to study the main quantities related to the energy deposition by protons, having clinical energies, in both liquid water and cortical bone. From Monte Carlo simulations with the Simulation of Energetic Ions and Clusters Through Solids (SEICS) code \cite{GarciaMolina2011,GarciaMolina2012SpringerScience}, we get the depth-dose curves, the penetration range and the WER in cortical bone, in order to compare them with the corresponding quantities for liquid water. To account for the condensed-phase nature of the studied materials, we use the Mermin Energy Loss Function -- Generalised Oscillator Strengths (MELF-GOS) method \cite{Abril1998, HerediaAvalos2005a} (based on the dielectric formalism) to calculate the electronic excitation spectrum of the target (encoded in its Energy Loss Function, ELF) as well as a multi-phase approach for cortical bone \cite{Limandri2014a}. This leads to stopping power values showing differences of the order of $10$--$15$\% between our calculations (based in the dielectric formalism and the MELF-GOS methodology) and tabulation by the International Commission on Radiation Units and Measurements (ICRU) \cite{ICRU49,Berger2005} below 200 keV for water and 1000 keV for bone. Despite the scarcity of experimental data of depth-dose curves in bone, our depth-dose simulation is in excellent agreement with Vasiliev \textit{et al.'s} experimental measurements \cite{vasiliev2010tissue} , even better than using ICRU data. In addition, our simulations of WER in cortical bone show small differences between using the MELF-GOS stopping power or the ICRU one. In spite of a fair agreement with recently reported WER values by Burin \textit{et al.} \cite{Burin2023}, the differences between both datasets are more visible.

This work is organized in the following sections. The methodological framework is  introduced in Section 2, including the calculation of the stopping quantities (Section 2.1), obtaining the energy loss function (ELF) of hydroxyapatite and cortical bone (Section 2.2), as well as a brief description of the simulation code (Section 2.3). The results found for the stopping quantities, the depth-dose curves and ranges for protons of several incident energies in both liquid water and cortical bone, as well as the WER values for bone are presented in Section 3. Finally, the main conclusions of this paper are summarized in Section 4.


\section{Theory and simulation methods}

\subsection{Calculation of the stopping quantities}

\hspace{0.82cm}The dielectric formalism  \cite{Fermi1940,Lindhard1954,Ritchie1957,Abril1998,HerediaAvalos2005a,Emfietzoglou2012} constitutes an appropiate framework to study the energy loss of fast protons in condensed-phase materials, as it includes the energy transfer by the projectile to the electronic excitations of the target accounting for its condensed-phase nature. 


Within this formalism, a projectile with mass $M$, atomic number $Z_1$, kinetic energy $T$ and charge state $q$ travelling through a medium, characterised by a dielectric function $\epsilon(k, \omega)$, will be stochastically slowed down with a stopping power $S_q$ (mean energy loss per unit path length) given by:
\begin{equation}
S_q (T) = \frac{Me^2}{\pi \hbar^2 T} \int_{E_{-}}^{E_{+}} \textrm{d}E \, E  \int_{k_{-}}^{k_{+}} \frac{\textrm{d}k}{k} [Z_1-\rho_q(k)]^2 \mbox{Im}\left[ \frac{-1}{\epsilon(k, E)} \right] 
\mbox{ , }
\label{eq:Sq}
\end{equation}
where $E$ and $\hbar k$ represent the energy and momentum of the target electronic excitations, respectively. 

The stochastic nature of the energy loss processes results in a distribution of values for the projectile energy loss processes, which is characterised by the so called energy loss straggling $\Omega_q^2$: 
\begin{equation}
    \Omega_q^2 (T) = \frac{Me^2}{\hbar^2 \pi T} \int_{E_{-}}^{E_{+}} \textrm{d} E \, E^2 \int_{k_{-}}^{k_{+}} \frac{\textrm{d}k}{k} [Z_1-\rho_q(k)]^2 \mbox{Im}\left[ \frac{-1}{\epsilon(k, E)} \right]
    \mbox{ . } 
\label{eq:Stg_q}
\end{equation}
In both equations \eref{eq:Sq} and \eref{eq:Stg_q}, $e$ denotes the elemental charge, whereas the electronic charge distribution of the projectiles is characterized by the Fourier transform of its electronic density, $\rho_q (k)$. 
The integration limits $E_+$ and $E_-$, as well as $k_+$ and $k_-$, are obtained by the fulfillment of the laws of conservation of energy and momentum \cite{Abril1998, deVera2018}.

Moreover, these formulae include the so-called energy-loss function (ELF) ${\rm Im}\left[ -1/\epsilon(k,E) \right]$. This quantity accounts for the probability of producing an excitation or ionization in the target with a certain energy and momentum, and it is calculated by using the MELF-GOS (Mermin Energy Loss Function -- Generalised Oscillator Strengths) method \cite{Abril1998, HerediaAvalos2005a}, as briefly described in what follows. 


The stopping power $S$ and the energy loss straggling $\Omega^2$ can be calculated as the weighted sum over $S_q$ and $\Omega_q^2$ for the different charge states $q$ of the projectile as it travels through the target capturing and losing electrons; $\phi_q$ represents the probability that the projectile has a charge state $q$ \cite{HerediaAvalos2005a, GarciaMolina2011}:
\begin{equation}
   S(T) = \sum_{q=0}^{Z_1} \phi_q (T) S_q(T) \mbox{ , } \qquad \Omega^2 (T) = \sum_{q=0}^{Z_1} \phi_q (T) \Omega_q^2 (T)    
    \mbox{ . } 
\label{eq:S-Stg}
\end{equation}

When the projectile produces an excitation in the target, the response of the target electrons will be different depending on their binding energies, so the ELF is split in two terms:
\begin{equation}
    {\rm Im}\left[ \frac{-1}{\epsilon(k,E)} \right] = {\rm Im}\left[ \frac{-1}{\epsilon(k,E)} \right]_{\mathrm {out}} + \sum_j \nu_j \sum_{nl} {\rm Im}\left[ \frac{-1}{\epsilon(k,E)} \right]_{nl}^{j} \, \mbox{.}
    \label{eq:ELF}
\end{equation}
The first term is due to the target outer-shell electrons, while the second one comes from the inner-shell electrons. 
In the last term, $\nu_j$ specifies the stoichiometry of the $j$th element in a compound target, while the indexes $nl$ refer to the ($n$,$l$) subshells of the inner electrons \cite{HerediaAvalos2005a}.

On the one hand, inner-shell electrons preserve their atomic character because of their large binding energies. Therefore, this kind of excitations are taken into account by using hydrogenic generalised oscillator strengths (GOS) for the K-, L- and M-shells to obtain the contribution to the ELF of the inner-shell electrons \cite{HerediaAvalos2005a}:
\begin{equation}
{\rm Im}\left[ \frac{-1}{\epsilon(k,E)}\right]_{nl}^{j} = \frac{2\pi^2 \hbar^2 e^2 {\cal N}}{m E} \,
\frac{{\rm d}f_{nl}^j(k, E)}{{\rm d} E} \, \Theta(E - E_{{\rm th},nl}^{j})\, \mbox{,}
\label{GOS}
\end{equation}
where $\cal{N}$ is the molecular density of the target, $m$ is the electron mass and $\rmd f_{nl}^j (k,E)/\rmd E$ is the hydrogenic GOS for the ($n$,$l$)-subshell of the $j$th atomic component, whose energy of ionization is $E_{{\rm th},nl}^{j}$. 

Due to the weak binding energy of the outer electrons, their excitations reflect the condensed-phase nature of the target and are better described from available data at the optical limit ($k=0$), using a weighted-sum of Mermin-type ELFs \cite{Mermin1970}, as described in Ref. \cite{HerediaAvalos2005a}: 
\begin{eqnarray}
\label{eq:ELFout}
{\rm Im}\left[ \frac{-1}{\epsilon(k=0,E)} \right]_{\rm out} & = & \sum_i \frac{A_i}{E_i^2} {\rm Im}\left[ \frac{-1}{\epsilon(E_i,\gamma_i;k=0,\hbar \omega)} \right]_{i} \label{eq:ELFi} \nonumber \\ 
 & = & \sum_i \Theta(E - E_{{\rm th},i}) \, \frac{A_i \, E \, \gamma_i}{(E^2-E_i^2)^2+(E \gamma_i)^2} \, \mbox{.} 
\end{eqnarray}
Here, $A_i$, $E_i$ and $\gamma_i$ are fitting parameters related with the intensities, positions and widths of the peaks appearing in the optical ELF. The function $\Theta(E - E_{{\rm th},i})$ denotes the step function of the transferred energy $\hbar \omega$, whose threshold energy is $E_{{\rm th},i}$. 


With the MELF-GOS method, the excitation spectrum of the condensed-phase target is appropriately described, automatically extending it to finite momentum transfers by means of the analytical properties of the Mermin dielectric function and the GOS, without using \textit{ad hoc} hypothesis for the $k$-dependence of the ELF. Actually, the MELF-GOS method has reproduced nicely \cite{Abril2010a} the experimental ELF of liquid water for $k \geq 0$ \cite{Watanabe1997,Hayashi2000}. 

\subsection{ELF of cortical bone}

In order to characterize the targets studied in this work, we apply the MELF-GOS method to liquid water and cortical bone. The former has been obtained in previous works by our group \cite{Emfietzoglou2008,Emfietzoglou2009,GarciaMolina2009,GarciaMolina2011}, by a fitting of the outer ELF to experimental data \cite{Watanabe1997, Hayashi2000} and obtaining the inner-shell contribution to the ELF from the GOS of the oxygen K-shell \cite{GarciaMolina2011}. 

Cortical bone, having a more complex composition \cite{ICRU49}, can be divided into a mineral contribution and an organic contribution \cite{Limandri2014a}. Therefore, to account for the inner-shells contribution to the ELF, we have to take into account the inner-shell electrons of all elements involved in both components. The mineral part, which is called calcium hydroxyapatite (HAp), contributes with inner-shell electrons from the K- and L-shells of calcium and phosphorus, as well as from the K-shell of oxygen \cite{Limandri2014a}. The last ones are also present in the inner-shells of the organic part of cortical bone, as well as the K- and L-shell electrons of phosphorus and sulfur, and the K-shell electrons of carbon and nitrogen. 

For the outer-shell electrons, we need to fit ${\rm Im} [-1/\epsilon (k, E)]_{\mathrm {out}}$ to the optical ELF of HAp or of organic bone. Unfortunately, there are no available experimental data for the former. However, \textit{ab initio} calculations of the optical properties of HAp were performed by applying density functional theory (DFT) within the local-density approximation \cite{Rulis2004} and fitted using equation \eref{eq:ELFout}, as explained in Ref. \cite{Limandri2014a}. 


The mass composition of cortical bone is 58\% of HAp and 42\% of organic material \cite{ICRU46}. Since the ELF of the former component is already known \cite{Limandri2014a}, it is needed to know the ELF of the second contribution. As no experimental measurements are available of the latter constituent of cortical bone, its optical-ELF can be described by means of a single-Drude function, whose parameters depend on the mean atomic number and the density of the target \cite{Tan2004}. This procedure has been applied to several organic targets \cite{Tan2004,deVera2013PRL,deVera2015}, obtaining  a nice agreement with available experimental measurements. 

Therefore, to know the optical ELF of the organic part of cortical bone the composition of this material is needed. This composition is obtained assuming that all the calcium was in the mineral part, so the HAp composition (with density $\rho =$ 1.85 g/cm$^3$) was subtracted from the cortical bone composition provided in Ref. \cite{ICRU49}. The density of the organic part of cortical bone was calculated taking into account the densities and mass compositions of both cortical bone and HAp \cite{Limandri2014a}. Useful information about each condensed-phase target is shown in \tref{tab-comp} and \tref{tab-prop-part}. The last two columns of \tref{tab-prop-part} gather the mean excitation energies for the different target materials, as obtained in the present work ($I_{\rm{MELF-GOS}}$) and those reported by ICRU Report 49 ($I_{\rm{ICRU 49}}$) \cite{ICRU49}. It is worth to notice that our calculated value of $79.4$ eV is closer to the updated ICRU value of $78$ eV \cite{Sigmund2009}.



\Table{\label{tab-comp}Composition of the targets used in this work.}
\br
&\centre{7}{Composition$^{\rm a}$ (\% mass)}\\
\ns
&\crule{7}\\
Target&\centre{1}{C}&\centre{1}{H}&\centre{1}{N}&\centre{1}{O}&\centre{1}{P}&\centre{1}{Ca}&\centre{1}{S}\\
\mr
Liquid water&$\00.00$&$11.19$&$\00.00$&$88.81$&$\00.00$&$\00.00$&$\00.00$\\
HAp Ca$_{10}$(PO$_{4}$)$_{6}$(OH)$_{2}$&$\00.00$&$\00.20$&$\00.00$&$41.41$&$18.50$&$39.89$&$\00.00$ \\
Organic part of cortical bone&$30.46$&$\09.74$&$\08.87$&$48.18$&$\01.62$&$\00.00$&$\01.13$\\
Cortical bone$^{\rm b, c}$&$14.43$&$\04.72$&$\04.20$&$44.61$&$10.50$&$20.99$&$\00.32$\\
\br
\end{tabular}
\item[] $^{\rm a}$ For simplicity, elements heavier than S with presence less than 1\% have been omitted in the composition.
\item[] $^{\rm b}$ The composition is the same as the one shown in the ICRU Report 49 \cite{ICRU49}.
\item[] $^{\rm c}$ Mass constitution of cortical bone is made of 58\% HAp and 42\% organic part \cite{Limandri2014a}.
\end{indented}
\end{table}

\Table{\label{tab-prop-part}Atomic number \textit{Z}, molecular mass \textit{A}, density $\rho$ and mean excitation energy \textit{I} of the targets considered in this work.}
\br
\ns
&&\centre{1}{$A$}&Density&\centre{1}{$I_{\rm{MELF-GOS}}$}&\centre{1}{$I_{\rm{ICRU 49}}$}\\
Target&$Z$&\centre{1}{(g/mol)}&(g/cm$^{3}$)&\centre{1}{(eV)}&\centre{1}{(eV)}\\
\mr
Liquid water&$\0\010$&$\0\018.02$&$1.00$&$\079.4$&$\075.0$\\
HAp Ca$_{10}$(PO$_{4}$)$_{6}$(OH)$_{2}$&$\0500$&$1008.83$&$3.22$&$159.5-162.5$&$140.2$\\
Organic part of cortical bone &$1633$&$2832.67$&$1.165$&$\074.6$&$~~~-$\\
Cortical bone&$3094$&$5934.94$&$1.85$&$114.9$&$106.4$\\
\br
\end{tabular}
\end{indented}
\end{table}

\subsection{Simulation of transport and energy deposition by a proton beam}
\label{sec:SEICS}

The SEICS code \cite{GarciaMolina2011,GarciaMolina2012SpringerScience} is used in this work to simulate the transport of therapeutical proton beams, with initial energies ranging from 50 MeV to 250 MeV, through liquid water and cortical bone. This simulation code is able to follow the trajectories of energetic ions through a condensed-phase target. It uses molecular dynamics methods to describe the motion of the projectile as well as Monte Carlo techniques to account for the stochastic nature of electronic energy loss, multiple elastic scattering, the electron charge-exchange between the proton and the medium, as well as the nuclear fragmentation of the projectile caused by non-elastic nuclear scattering processes \cite{GarciaMolina2011,GarciaMolina2012WorldScientific,GarciaMolina2014, deVera2018}. The code uses a (relativistic) variant of the velocity Verlet's algorithm to calculate the new position $\vec r$ and velocity $\vec v$ of a projectile of mass $M$ after a time $\Delta t$  \cite{Allen1989}:
\begin{eqnarray}
\vec{r}_{i+1} & = & \vec{r}_i + \vec{v}_i \Delta t+\frac{\vec{F}_i}{2M}\mbox{ }(\Delta t)^2 \mbox{ }\left[ 1-\left(\frac{v_i}{c}\right)^2 \right]^{3/2} \mbox{ }{\rm ,} \label{eq:r-Verlet-rel} \\
\vec{v}_{i+1} & = & \vec{v}_i+\frac{\vec{F}_i+\vec{F}_{i+1}}{2M}\mbox{ }\Delta t \mbox{ }\left[1-\left(\frac{v_i}{c}\right)^2 \right]^{3/2} \mbox{ }{\rm ,}
\label{eq:v-Verlet-rel}
\end{eqnarray}
where $c$ is the velocity of light, and the subscripts $i$ and $i+1$ refer to successive time steps separated by a time interval $\Delta t$.  

In equations \eref{eq:r-Verlet-rel} and \eref{eq:v-Verlet-rel}, the factor in brackets is an \textit{ad hoc} correction which takes into account the relativistic velocity of the projectile, whereas $\vec F$ is the electronic stopping force \cite{deVera2018}. This force, which is mainly caused by inelastic collisions with the target electrons, acts on the proton with a charge state $q$. Because of its stochastic nature, it is randomly sampled by a Gaussian distribution whose mean value is the stopping power $S_q$ and whose variance is described in terms of the energy loss straggling $\Omega_q^2$, $\sigma = \sqrt{\Omega_q^2/\Delta s}$, where $\Delta s$ is the distance travelled by the proton in a time $\Delta t$, i.e., $\Delta s = v \Delta t$. Therefore, the expression of the electronic stopping force is \cite{Box1958}:
\begin{equation}
\vec{F} = -\left[ S_q + \frac{\Omega_q}{\sqrt{\Delta s}} \sqrt{-2\ln{R_1}}\cos{(2 \pi R_2)} \right]  \hat{v} \mbox{
}{\rm ,} \label{eq:A6}
\end{equation}
where $R_1$ and $R_2$ are random numbers uniformly distributed between 0 and 1, and $\hat{v}$ is the unit velocity vector. Both energy loss quantities, $S_q$ and $\Omega_q^2$, are calculated using equations \eref{eq:Sq} and \eref{eq:Stg_q}, with the ELF obtained through the MELF-GOS method.


The simulation code SEICS considers not only the elastic interactions between the projectile and the target nuclei but also the nuclear fragmentation reactions \cite{deVera2018}. These are needed to properly characterize the depth-dose curves, accounting for the interactions between the primary protons and the ones of the atoms of the medium. These interactions involve the excitation of the target nucleus, its fragmentation, the emission of secondary energetic particles (such as neutrons, photons, secondary protons or heavier particles) and the relaxation of the residual nucleus. 

Nuclear fragmentation reactions are approximately included in the SEICS code by removing the primary protons from the beam according to their total nuclear fragmentation cross section, locally depositing a part of their energy \cite{deVera2018}. The information related to nuclear fragmentation cross sections is obtained from ICRU Report 63 \cite{ICRU63}, a validated and comprehensive cross-section compilation based on theoretical models and available experimental data. The fragmentation mean free path $\lambda_{\rm fr}(v)$ for protons, which depends on their velocity $v$, is given by:
\begin{equation}
\lambda_{\rm fr}(v) = \frac{A}{N_{\rm A} \, \rho \, \sigma_{\rm fr}(v)} {\rm ,} \label{eq:FrMFP}
\end{equation}
where $A$ and $\rho$ are, respectively, the mass number and density of the target, whereas $N_{\rm{A}}$ refers to Avogadro's number. In addition, $\sigma_{\rm fr}(v)$ is the microscopic fragmentation cross section, which is calculated as the weighted sum of the fragmentation cross sections for the elements that constitute the compound target \cite{deVera2018}. 

After a distance $\rmd s$, the number of surviving particles of the beam is $N(s+{\rm d}s) = N(s){\rm e}^{-{\rm d}s/\lambda_{\rm fr}}$, showing an exponential behaviour. In order to take this into account, a random number is sampled in each step $\rmd s$. If it is equal of less than the probability of fragmentation, namely, $1-N(s+{\rm d}s)/N(s) \leq 1$, then the primary proton disappears and a part of its energy is locally deposited. 

Because of their low stopping cross sections, the energy of secondary neutrals (photons and neutrons) escapes from the studied volume and this fraction of energy is disregarded. Knowing this, as well as the fact that secondary protons have long ranges compared with those of heavier particles, the treatment is normally simplified by tracking only secondary protons and depositing locally the energy of heavier secondary particles \cite{Medin1997}. Nevertheless, in the SEICS code the energy of both protons and heavier particles is locally deposited for simplicity \cite{deVera2018}. It will be seen later how this approximation is enough for our purposes.


\section{Results and discussion}

\begin{figure}[t]
 \centering
 \includegraphics[width=\textwidth]{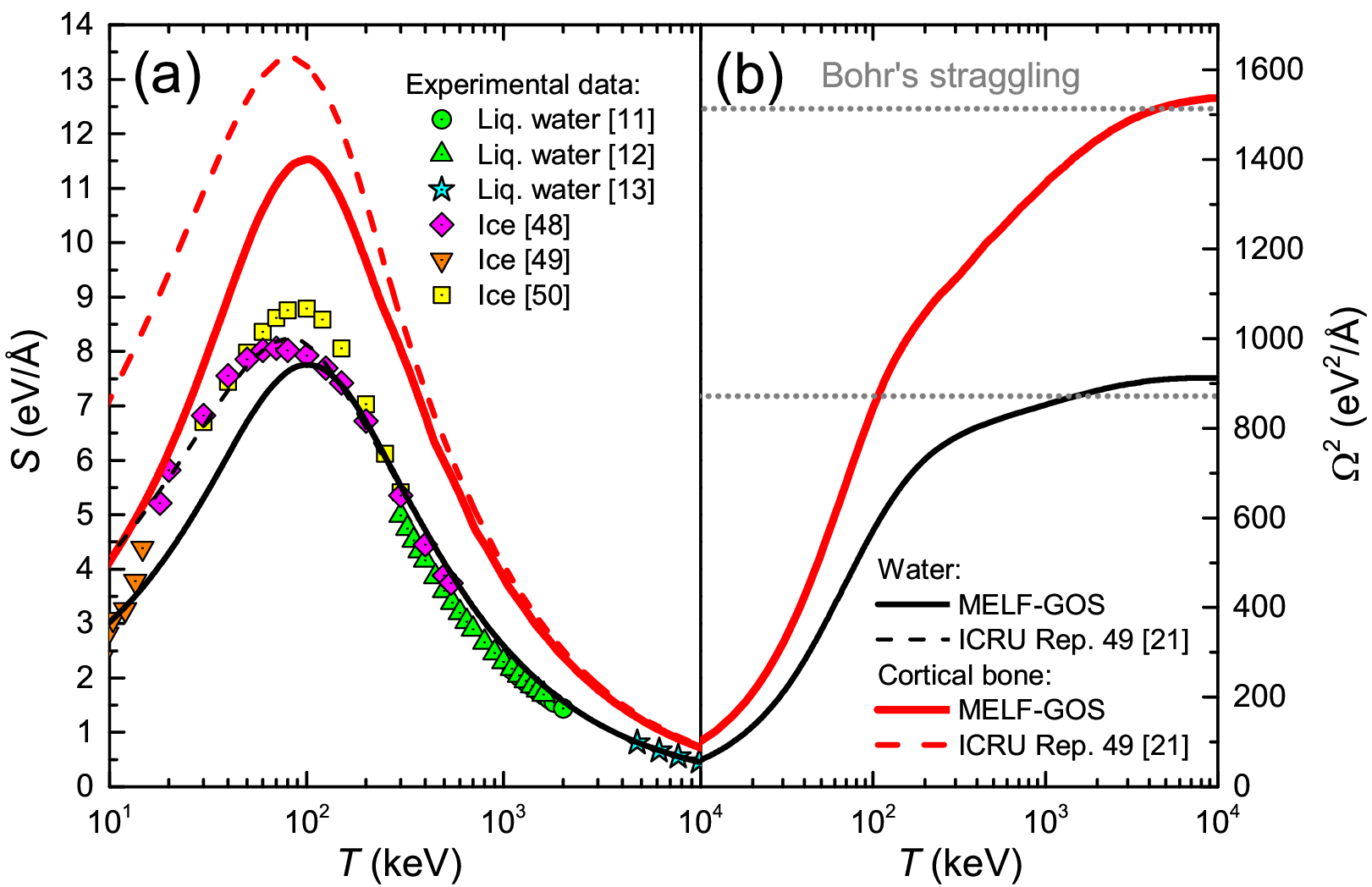}
  \caption{Stopping power (a) and energy loss straggling (b) for protons in liquid water and cortical bone, as a function of the beam energy. Lines represent calculations (from the current work and from ICRU \cite{ICRU49,Berger2005}), while symbols correspond to experimental data for liquid water and ice, as described in the legend. 
  }
\label{fig:spstr}
\end{figure}

The stopping power and energy-loss straggling of liquid water and cortical bone for protons are calculated using equations \eref{eq:Sq} and \eref{eq:Stg_q}, considering the equilibrium charge fractions $\phi_q$ of the projectile, which depend on its energy and the target. 
We have obtained the values of $\phi_q$ in liquid water and in cortical bone from the CasP code \cite{Schiwietz2001}, which is based on a parameterisation to experimental data.
\Fref{fig:spstr}(a) shows the stopping powers of liquid water and of cortical bone as a function of the incident proton energy $T$. Solid lines correspond to the results obtained in the current work by means of the MELF-GOS method; calculations for cortical bone are shown by a red thick solid line, while the ones for liquid water are represented by a black thin solid line. For comparison, the stopping powers provided by the ICRU Report 49 for both targets are depicted by dashed lines \cite{ICRU49,Berger2005}. Available experimental data for ice \cite{Wenzel1952,Andrews1977,Bauer1994} and liquid water \cite{Shimizu2009,Shimizu2010,Siiskonen2011} are depicted by symbols, while for cortical bone there are no measured data. 



Our results for liquid water excellently agree with the experimental data for liquid water at high energies (star symbols) \cite{Siiskonen2011}, while slight differences are found with respect to the experiments in liquid water at intermediate energies (triangle and circle symbols) \cite{Shimizu2009,Shimizu2010}. However, it should be taken into account that the latter experiments were performed in liquid water jets, and the experimentally measured proton energy spectra had to be interpreted by means of Monte Carlo simulations in order to infer the stopping power \cite{Shimizu2009,Shimizu2010}; such analysis was done on the basis of a scaling of the stopping power curve given by the SRIM code \cite{SRIM2013}. Garcia-Molina \textit{et al.} \cite{GarciaMolina2013NIMB} showed how the stopping power calculated from the dielectric formalism, without any \textit{ad hoc} scaling, could also reproduce in detail the measured spectra in water and other cylindrical targets. Thus, the experimentally derived stopping powers of liquid water at intermediate proton energy should be taken with some care. Our calculations also closely follow the experimental stopping power for ice \cite{Wenzel1952,Bauer1994} down to $\sim 200$ keV, underestimating it below this energy. Such differences could be attributed to the limitations of the dielectric formalism at low proton energies and/or to phase differences between ice and liquid water.

Our calculations agree well with ICRU data \cite{ICRU49,Berger2005} for both targets at high energies. However, the latter systematically overestimate our results at energies $\lesssim 200$ keV for liquid water and $\lesssim 1000$ keV for cortical bone. It should be taken into account that ICRU estimates take into account the experimental information available for each target material; as for cortical bone there are no measurements, it is natural that ICRU values present larger deviations with respect to the present calculations based on the dielectric formalism.

\Fref{fig:spstr}(b) depicts the calculated energy-loss straggling of both targets for protons as a function of the beam energy. Again, the red thick solid line corresponds to cortical bone, whereas the black thin solid one is for liquid water. Due to the fact that there are no available data to compare our results with, the asymptotic high energy value given by the Bohr's straggling formula is included as dotted lines, as a way to contrast our calculations. This quantity has been calculated as \cite{Sigmund2006}:
\begin{equation}
\Omega_{\rm Bohr}^2 = 4\pi e^4 Z_1^2 Z_2 {\cal N} {\rm ,}
\label{eq:BohrStg}
\end{equation}
where $Z_1$ is the atomic number of the projectile, whereas $Z_2$ and ${\cal N}$ are the atomic number and the molecular density of the target. The calculated stragglings converge reasonably well to the corresponding asymptotic Bohr's values.



\begin{figure}[t]
 \centering
 \includegraphics[width=\textwidth]{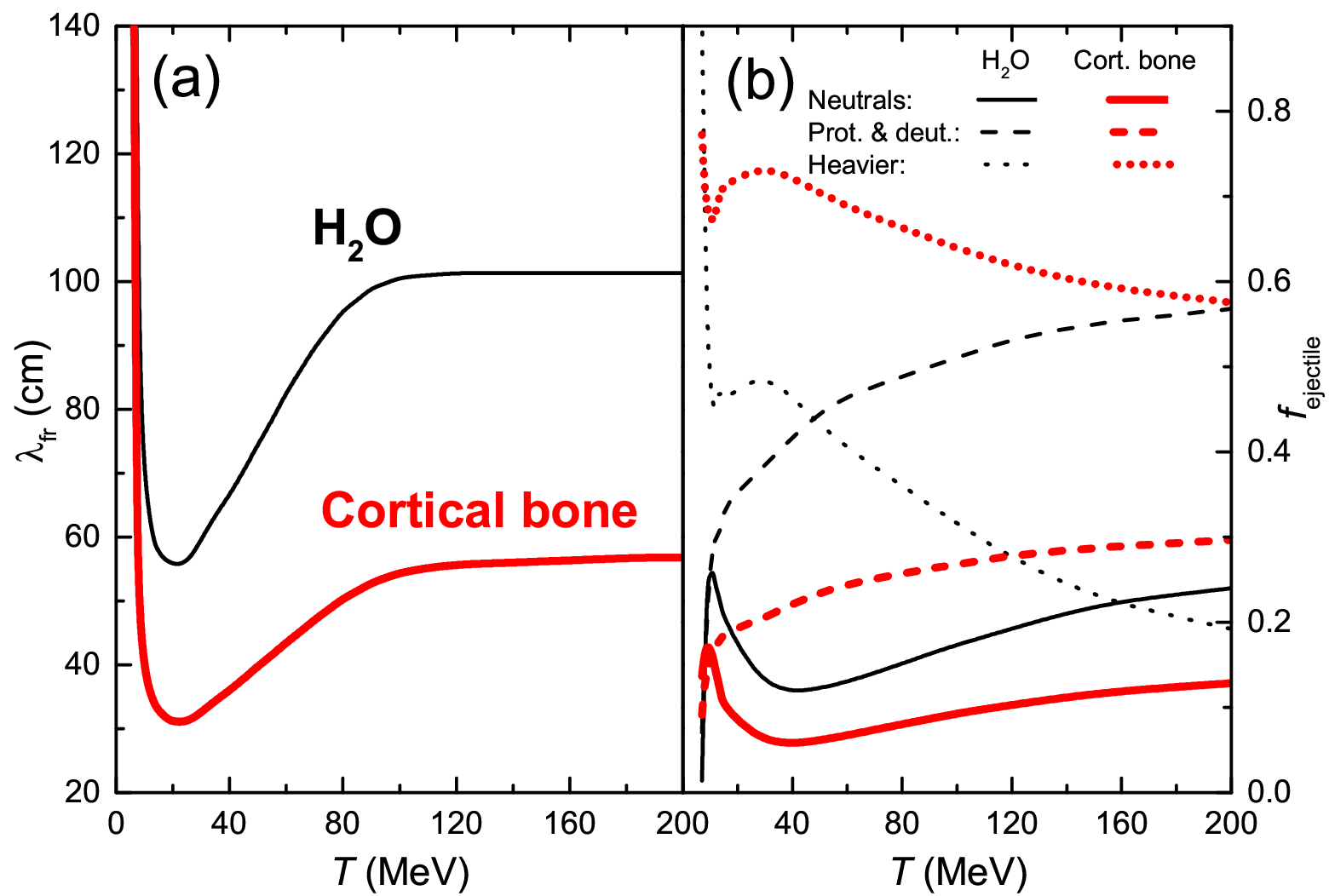}
  \caption{(a) Nuclear fragmentation mean free paths for protons in liquid water and cortical bone, and (b) fractions of proton energy delivered to different ejectiles, as a function of proton beam energy. All the quantities are derived from data by ICRU \cite{ICRU63}, as explained in the text. 
  }
\label{fig:lfrag}
\end{figure}

It is worth to notice that the calculated energy loss quantities for cortical bone are always larger than for liquid water, which should be taken into account if detailed simulations of energetic proton beams interacting with the human body for oncological purposes (i.e., protontherapy) are needed. 

\begin{figure}[t]
 \centering
 \includegraphics[width=\textwidth]{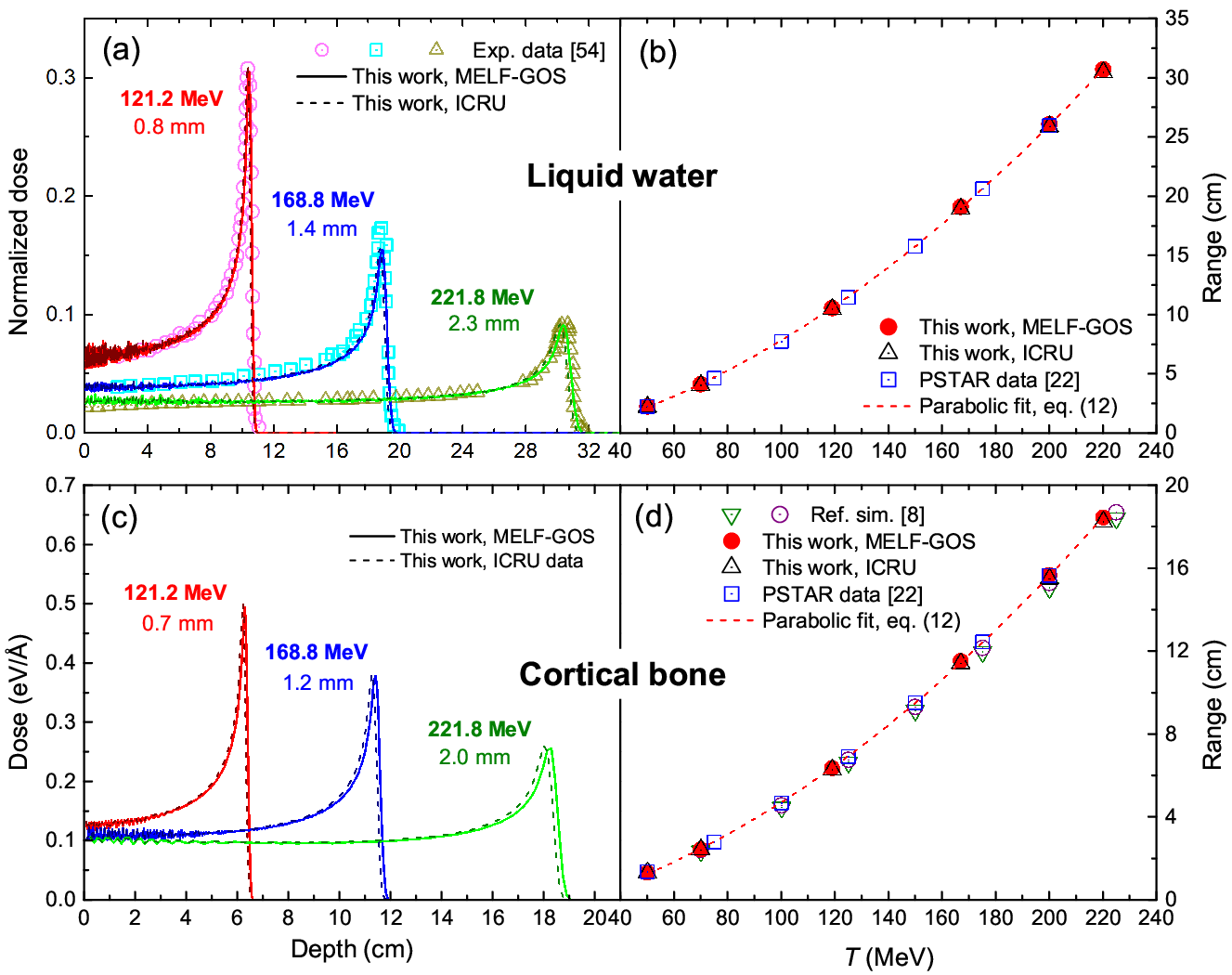}
  \caption{Panels (a) and (c) show simulated depth-dose curves (lines) for proton beams of different initial energies (given in the labels) in liquid water and cortical bone, respectively. Solid lines correspond to simulations using MELF-GOS stopping quantities, while dashed lines use ICRU stopping powers. Symbols depict experimental results for liquid water \cite{Zhang2011}. The labels below the initial energies inform about the difference in proton ranges obtained in each case, when using MELF-GOS or ICRU stopping powers. Panels (b) and (d) show, respectively, the simulated proton ranges as a function of the initial beam energy, extracted from the different simulations, and compared to results from Ref. \cite{Burin2023} and from NIST's PSTAR database \cite{Berger2005}. Red dashed lines give the best fits to the MELF-GOS results by means of Eq. (\ref{eq:RangeFit}).
  }
\label{fig:BraggRanges}
\end{figure}

The previous data are necessary to simulate the propagation of proton beams in liquid water and cortical bone, together with the corresponding nuclear fragmentation information for each target. \Fref{fig:lfrag}(a) shows the nuclear fragmentation mean free paths for protons in liquid water (dashed black line) and cortical bone (solid red line), as a function of proton energy, obtained from the data from ICRU Report 63 \cite{ICRU63}, as explained in \sref{sec:SEICS}. Cortical bone presents a rather shorter fragmentation mean free path, due to its larger density and the heavier elements present in its composition (see \tref{tab-comp} and \tref{tab-prop-part}). The fraction $f_\textrm{\tiny ejectile}$ of proton energy transferred to different ejectiles (neutrals, protons and deuterons, and heavier particles) is depicted in \fref{fig:lfrag}(b), for both liquid water (thin black lines) and cortical bone (thick red lines), as a function of the primary proton energy. In cortical bone, lower fractions of neutral projectiles and secondary protons are produced with respect to water, whereas the amount of heavier ejectiles is much larger. This will imply that simulations using the SEICS code (which locally deposits the residual energy of both short-range heavier ions --correctly-- and long-range secondary protons --incorrectly--) will be more accurate for cortical bone than for liquid water, which are excellent, as will be seen when comparing simulated depth-dose curves with available experimental data in \Fref{fig:BraggRanges}(a).

\begin{figure}[t]
 \centering
 \includegraphics[width=0.9\textwidth]{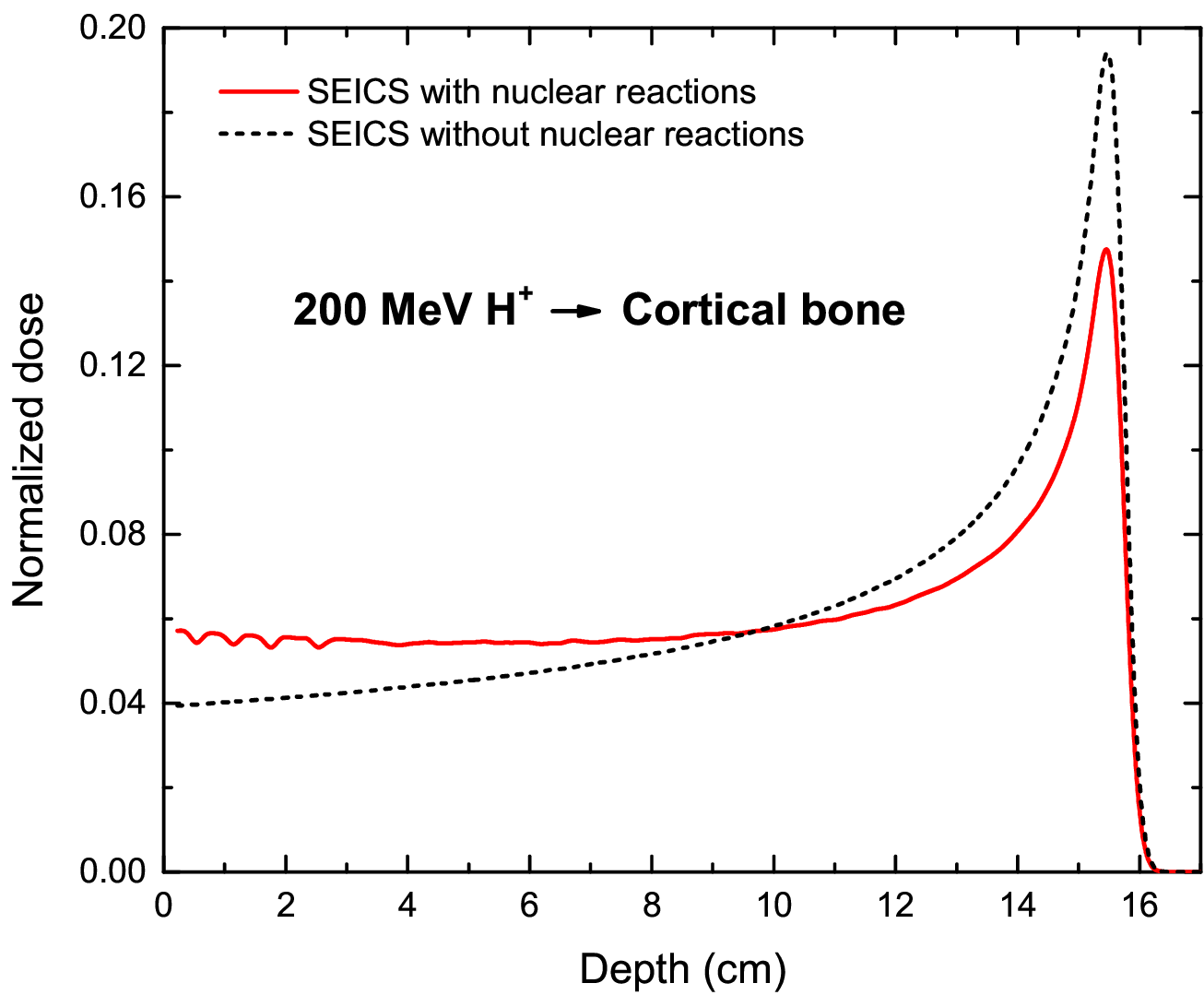}
  \caption{Effect of nuclear fragmentation reactions on the depth-dose curve of $200$ MeV protons in cortical bone. The dashed line corresponds to a simulation in which nuclear fragmentation reactions are not considered, whereas the solid curve shows the result of a full simulation, where nuclear reactions are included as described in the text.
  }
\label{fig:FrafInDose}
\end{figure}

As an illustration of the influence of nuclear fragmentation reactions, \fref{fig:FrafInDose} depicts the simulated depth-dose curves for 200 MeV protons in cortical bone, including (solid line) and disregarding (dashed line) the nuclear reactions. The curves have been normalized, so as the areas under the curves have a value of one. Nuclear fragmentation reactions tend to decrease the dose around the Bragg peak maximum, due to the loss of primary ions, whose slowing down is more intense around these depths. By contrast, the dose is increased in the beam entrance region, where primary protons present larger residual energies when fragmenting, as the SEICS code deposits locally this energy. 

In Ref. \cite{deVera2018}, depth-dose curves for proton beams in liquid water, as simulated by the SEICS code, were compared with available experimental data \cite{Zhang2011} for several initial energies. Such comparison is shown in \fref{fig:BraggRanges}(a), together with current results for the same beam energies in cortical bone in \fref{fig:BraggRanges}(c). Experimental depth-dose curves have also been scaled to unit area. There are no experimental depth-dose curves in bone for these energies, but comparison with measured data will be reported later. For liquid water, the agreement between simulated and experimental depth-dose curves is excellent. For the largest energy shown (221.8 MeV), the simulated dose is slightly overestimated in the entrance region, due to the lower reliability of the local-energy deposition scheme used for secondary protons at these depths. Still, the agreement is very good. 

In both \fref{fig:BraggRanges}(b) and (d), simulated results are shown feeding the SEICS code either with the currently calculated stopping powers and energy loss stragglings (solid lines) or with ICRU stopping powers, supplemented with our calculated stragglings (dashed lines). For liquid water, the shift of the depth-dose curves is difficult to appreciate in the figure, but the differences in the determined proton ranges (estimated as the depths were the relative dose falls to $80$\% after the maximum) is shown by labels, in units of mm, below each energy label. Such differences grow with the initial beam energy, going up to $2.3$ mm at $221.8$ MeV. Similar absolute differences in the proton ranges are found for cortical bone using the present or the ICRU slowing down data. However, it should be noted that, since the ranges in cortical bone are considerably shorter than in liquid water, the relative differences are more significant. \Fref{fig:BraggRanges}(b) and \fref{fig:BraggRanges}(d) show, respectively, the proton ranges in liquid water and cortical bone as a function of the beam energy. Full circles represent SEICS simulations using our calculated stopping powers and energy loss stragglings, while up triangles correspond to simulations using ICRU stopping powers. The energy-range simulations are compared with results given for these targets by NIST's PSTAR database \cite{Berger2005}, as well as with other simulated results for cortical bone \cite{Burin2023}. While our simulations seem to agree well with PSTAR data, differences with the results of Ref. \cite{Burin2023} are more visible.

Our simulated results for the proton range (in cm) are very well reproduced by the following energy--range relation:
\begin{equation}
{\rm Range~ (cm)} = a + b \cdot T~ ({\rm MeV}) + c \cdot T~ ({\rm MeV})^2 {\rm ,}
\label{eq:RangeFit}
\end{equation}
where $T$ is the initial proton beam energy (in MeV), whereas  $a$, $b$ and $c$ are fitting parameters, whose value appear in \tref{tab-fit}. Red dashed lines in \fref{fig:BraggRanges}(b) and \fref{fig:BraggRanges}(d) denote the best fits of this equation to the ranges of protons (for energies $50$--$220$ MeV) in liquid water and cortical bone, respectively, determined by simulations using the MELF-GOS stopping quantities. 

\Table{\label{tab-fit} Parameters used in the parabola fitting of the energy-range relationship for liquid water and cortical bone.}
\br
Material&$a$ (MeV)&$b$ (cm/MeV)&$c$ (cm/MeV$^2$)\\
\mr
Liquid water&$-1.23341$&$4.536 \cdot 10^{-2}$&$4.54404 \cdot 10^{-4}$\\
Cortical bone&$-0.85881$&$2.932\cdot 10^{-2}$&$2.6587\cdot 10^{-4}$\\
\br
\endTable

\begin{figure}[t]
 \centering
 \includegraphics[width=\textwidth]{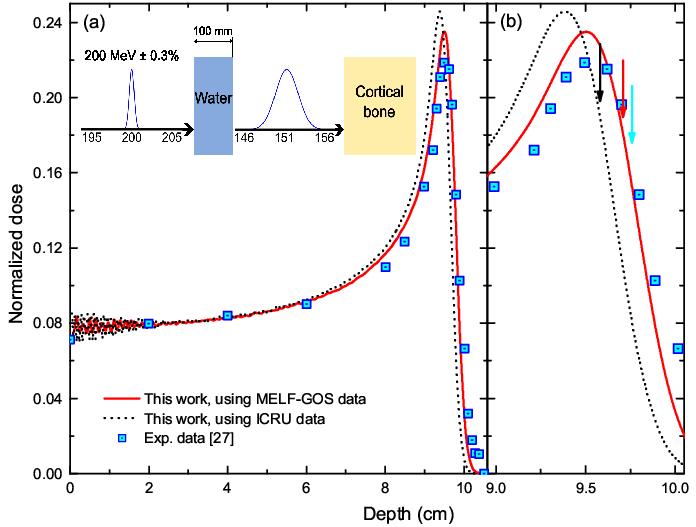}
  \caption{(a) Depth-dose curve in cortical bone, for an initial $200$ MeV proton beam, previously passed through a $100$ mm-wide target of liquid water. Symbols depict the results of experimental measurements in a cortical bone phantom \cite{vasiliev2010tissue}, while red solid (black dashed) line corresponds to a SEICS simulation using the MELF-GOS (ICRU) stopping power. All curves are normalized to unit area. The inset sketches the experimental setup, in which a $200$ MeV $\pm \, 0.3$ \% initial proton beam traverses a liquid water target, resulting in its energy spectrum widened before entering the cortical bone target. (b) Detail of the depth-dose curves around the maximum dose, where arrows indicate the location of the penetration range for each curve, estimated as the depth at which the dose falls to $80$ \% after the maximum.
  }
\label{fig:ExpBragg}
\end{figure}

The experimental information both on the stopping power and on the depth-dose curves for proton beams in cortical bone is very scarce, despite the importance of this target in radiotherapy treatments. The only measured depth-dose curve we are aware of is that of $200$ MeV protons interacting with a cortical bone phantom material reported by Vasiliev \textit{et al.} \cite{vasiliev2010tissue}. In the experimental setup, the $200$ MeV beam, having an energy width of $\pm 0.3$\% , was first passed through a liquid water energy degrader $100$ mm in depth. The resulting, further energy widened, beam then bombarded the cortical bone phantom. This setup is sketched in the inset of Fig. \ref{fig:ExpBragg}(a).

The SEICS code allows to introduce an energy spectrum to the incident beam, as well as to measure the energy spectrum of the protons leaving the target slab. The spectra of the protons entering and leaving the liquid water target of $100$ mm width are depicted in the inset of Fig. \ref{fig:ExpBragg}(a). The energy loss straggling of liquid water makes the beam to increase its energy width from the initial $\pm 0.3$\% to $\pm 0.93$\% when exiting the liquid water slab. This latter value has been then used to simulate the proton propagation in the cortical bone target. The red solid line in Fig. \ref{fig:ExpBragg}(a) corresponds to the simulation using the stopping powers and energy loss stragglings obtained in this work (both for liquid water and cortical bone), while the black dashed line uses the ICRU stopping powers, complemented with the current energy loss straggling values. As can be clearly seen, MELF-GOS slowing down quantities yield a simulated depth-dose curve which almost perfectly agrees with the experimentally measured data from Ref. \cite{vasiliev2010tissue}, except for a slight overestimation of the dose around the Bragg peak maximum (again, results have been normalized, so as the areas under the depth-dose curves have a value of one). Such a good result (particularly, the excellent reproduction of the depth for maximum dose) has been obtained without the need of slight adjustments in the nominal initial beam energy, which sometimes are needed due to the initial energy uncertainties \cite{deVera2018}. In contrast, the simulation using the ICRU stopping power (for both target materials) results in a Bragg curve maximum slightly shifted to shallower depths, together with a somewhat more intense dose around the maximum, as easily observed in the figure. 
Figure \ref{fig:ExpBragg}(b) focuses around the maximum of the depth-dose curve, in order to better appreciate the scale of the range differences. The range obtained from each curve has been marked by arrows. Clearly, the range obtained using current stopping quantities ($\sim 9.71$ cm) is in closer agreement to the experimental value ($\sim 9.75$ cm) than that obtained from simulations using ICRU stopping power ($\sim 9.58$ cm).
The range variation associated with the use of one set of stopping powers or the other is of $\sim 1.3$ mm. This comparison supports the accuracy of the slowing down quantities calculated by means of the MELF-GOS method, as well as the good performance of the SEICS simulation code.

\begin{figure}[t]
 \centering
 \includegraphics[width=\textwidth]{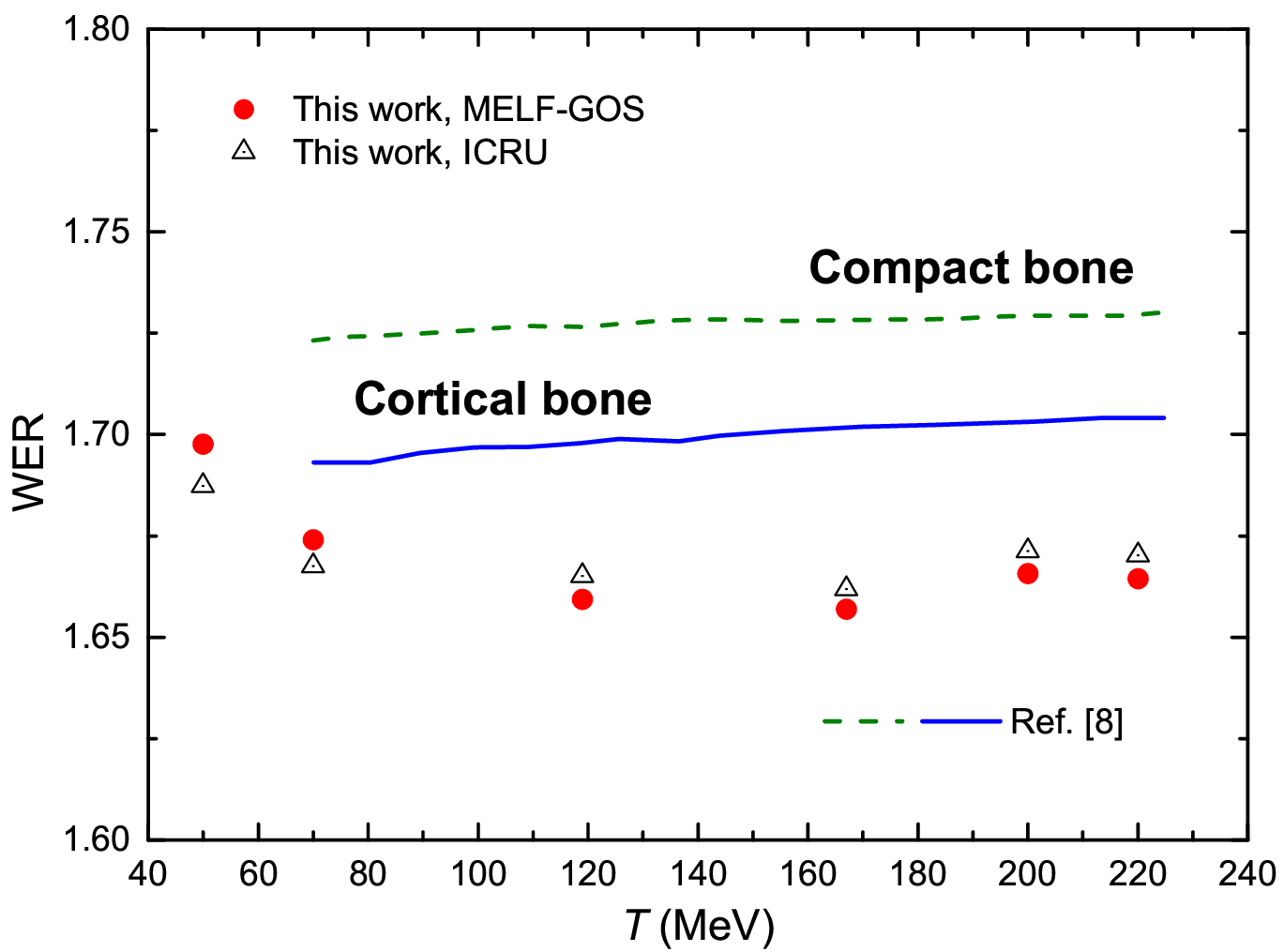}
  \caption{Water equivalent ratio of cortical bone as a function of proton beam energy. Solid and dashed lines depict, respectively, previous simulations performed by Burin \textit{et al.} \cite{Burin2023} for cortical and compact bone targets, while full (open) symbols correspond to current simulated results for cortical bone, using the MELF-GOS (ICRU) stopping power.
  }
\label{fig:WER}
\end{figure}

Simulated Bragg curves for proton beams in cortical bone can be used to obtain the energy-range relation (Fig. \ref{fig:BraggRanges}(d)) and, from it, to calculate the water equivalent ratio (WER) for this material, an important quantity in protontherapy treatment planning. The WER is obtained as the ratio of the cortical bone range and the liquid water range for a given proton beam energy. WER values were recently reported for cortical and compact bone by \cite{Burin2023} on the basis of Monte Carlo simulations by means of the MCNP6.1 and TOPAS codes. These data are shown as lines in Fig. \ref{fig:WER}, together with our calculations based on the dielectric formalism (ICRU) calculations and the SEICS code, depicted by full (open) symbols. In general terms, Burin \textit{et al.}'s and our results are compatible, being both close to $1.70$ in the energy range from $50$ to $220$ MeV. The results based on SEICS, using either the MELF-GOS or the ICRU slowing down quantities, present a similar behaviour with the beam energy, slowly decreasing from $\sim 1.70$ at $50$ MeV to $\sim 1.65$ from around $120$ MeV, with small variations among them. However, the evolution of the WER curve obtained by Burin \textit{et al.} \cite{Burin2023} is different, presenting a slight monotonic increase in the studied energy range, being always very close to $1.70$. Further simulations, using other Monte Carlo codes and different stopping power input data, would be needed to more accurately determine the WER of protons in cortical bone, particularly at the energies $> 100$ MeV commonly employed in protontherapy.

\section{Summary and conclusions}


Cortical bone is a very relevant material, almost unavoidably encountered by energetic proton beams during cancer treatment by means of protontherapy. However, despite the importance of  the presence of this material in treatment plans, which will introduce uncertainties in the penetration range of proton beams in the human body, there is still a current lack of accurate information on its energy loss quantities, such as the stopping power and the energy loss straggling. In fact, the assessment of the water equivalent ratio (WER) of cortical bone has only very recently been undertaken by means of general purpose Monte Carlo codes such as TOPAS or MCNP \cite{Burin2023}, though using simple approximations to the stopping power, such as the Bragg's rule or ICRU estimates \cite{ICRU49,Berger2005}. The former introduces uncertainties related to the chemical bonding and condensed-phase state of the target, while the accuracy of the latter relies on the availability of measured data, which is lacking in this case. 

The dielectric formalism represents a convenient theoretical method to calculate accurate proton stopping powers and energy loss stragglings in condensed-phase materials.
However, its application requires the knowledge of the target ELF, which is difficult to determine for cortical bone, given its multi-phase nature. In Ref. \cite{Limandri2014a}, the results of DFT calculations of the ELF of calcium hydroxyapatite (the mineral phase of cortical bone) \cite{Rulis2004} were combined with an estimate of the ELF of the organic part of cortical bone, obtained from a parametric model to predict the ELF of organic materials \cite{Tan2004,deVera2013PRL}. Experimental measurements of the stopping power of proton and helium ions in hydroxyapatite performed by \cite{Limandri2014a} demonstrated the good accuracy of the dielectric formalism, together with the MELF-GOS method to describe the target ELF. An appropriate combination of the stopping powers for the mineral and organic parts of cortical bone also predicted its stopping power in good agreement with the scarce available information for this target \cite{Limandri2014a}.

Still, a full validation of the calculated slowing down quantities requires the reproduction of other experiments, of relevance for protontherapy, by means of radiation transport simulations. We sought to simulate proton depth-dose curves in cortical bone by means of the SEICS code, implementing our calculated slowing down quantities. However, to the best of our knowledge, there is only one experimental measurement \cite{vasiliev2010tissue}, whose data are, nonetheless, very useful: a $200$ MeV proton beam (with a well characterized energy dispersion) first traversed a liquid water energy degrader, to then bombard cortical bone. Thus this experiment tests the stopping quantities of both liquid water and cortical bone.

As shown in the present work, our calculated stopping power of cortical bone presents significant deviations from the ICRU49 estimate \cite{ICRU49,Berger2005} for energies below $1000$ keV, while these differences only appear below $200$ keV for liquid water (a better characterized target, for which there exist experimental stopping power data, even though scarce).

Such deviations are significant enough to produce visible differences in the simulated depth-dose curves in cortical bone: while our calculated stopping quantities lead to a simulated Bragg curve in very good agreement with the measurements by Vasiliev \textit{et al.} \cite{vasiliev2010tissue}, the use of ICRU49's stopping power yields a depth-dose curve with the peak position slightly off, and with a somewhat worse general agreement with experiment. Thus, these results demonstrate the high accuracy of our calculated slowing down quantities, and their visible impact in clinically relevant magnitudes for both important constituents of human body, namely, liquid water and cortical bone.

Provided the excellent agreement of our calculated energy loss quantities, we used them in our SEICS simulation code to study the proton range in liquid water and cortical bone as a function of the beam initial energy. The ratio of the ranges in cortical bone and liquid water can then be used to obtain the water equivalent ratio (WER) for cortical bone as a function of the beam energy. The differences in the simulated penetration range of protons in these materials, when using our current stopping powers or those given by ICRU \cite{ICRU49,Berger2005} progressively grow with the beam energy. Range differences are around $0.7$--$0.8$ mm for $121.2$ MeV protons, while they increase to $2.0$--$2.3$ mm for $221.8$ MeV. In relative terms, such differences are more significant for cortical bone than for liquid water, as absolute ranges are rather shorter in the former. The resulting ranges obtained by using the currently calculated stopping powers and energy loss stragglings are well fitted by parabolic functions in the energy range studied. 

The differences in penetration range reflect in differences in the WER values for cortical bone. Simulated WER with SEICS, using either our calculated stopping quantities or those given by ICRU49, present slight deviations, even though their evolution with beam energy is similar. They decrease from $\sim 1.70$ at $50$ MeV (a value similar to that predicted by Burin \textit{et al.} \cite{Burin2023}) to $\sim 1.65$ at energies around $120$--$160$ MeV, with a very slight increase afterwards. In contrast, the WER derived in Ref. \cite{Burin2023}, even though close, presents a slightly different evolution with energy, with a very small but constant increase, being always around $1.70$.

As current treatment plannings typically require millimetric precision, and given the fact that cortical bone is difficult to avoid in many clinical situations, the observed differences in the proton ranges and WER values may contribute to undesired uncertainties during protontherapy treatments. It is clear then the importance to know the slowing down quantities (especially the stopping power, but also the energy loss straggling) of protons in cortical bone, at least as accurately as they are known for liquid water. In this work we have presented theoretical calculations and simulations, based on the dielectric formalism, which, by comparison with the experiments performed with calcium hydroxyapatite \cite{Limandri2014a} and the measured depth-dose curve for cortical bone phantom \cite{vasiliev2010tissue}, seem to offer a very good accuracy. Experimental determinations of the stopping power of protons in cortical bone would be very useful to further check the presented results, as well as to better understand the differences in WER as determined in the present work and those recently obtained in Ref. \cite{Burin2023}.

\vspace{10pt}

The datasets generated during the current study and simulation code are available from the corresponding author on reasonable request.

\ack

This work is part of the R\&D project no. PID2021-122866NB-I00 funded by the Spanish Ministerio de Ciencia e Innovación (MCIN/AEI/10.13039/501100011033/) and by the European Regional Development Fund (“ERDF A way to make Europe”), as well as of the R\&D project no. 22081/PI/22 funded by the Autonomous Community of the Region of Murcia through the call ``Projects for the development of scientific and technical research by competitive groups'', included in the Regional Program for the Promotion of Scientific and Technical Research (Action Plan 2022) of the Fundación Séneca – Agencia de Ciencia y Tecnología de la Región de Murcia.

\section*{References}


\bibliography{library}

\providecommand{\newblock}{}
\begin{thebibliography}{10}
\expandafter\ifx\csname url\endcsname\relax
  \def\url#1{{\tt #1}}\fi
\expandafter\ifx\csname urlprefix\endcsname\relax\def\urlprefix{URL }\fi
\providecommand{\eprint}[2][]{\url{#2}}

\bibitem{Kraft1990}
Kraft G 1990 {\em Strahlentherapie und Onkologie\/} {\bf 166} 10

\bibitem{Elsasser2009}
Els{\"{a}}sser T, Gemmel A, Scholz M, Schardt D and Kr{\"{a}}mer M 2009 {\em
  Physics in Medicine and Biology\/} {\bf 54} 101--106 ISSN 0031-9155
  \urlprefix\url{http://www.ncbi.nlm.nih.gov/pubmed/19287080}

\bibitem{Schardt2010}
Schardt D, Els{\"{a}}sser T and Schulz-Ertner D 2010 {\em Reviews of Modern
  Physics\/} {\bf 82} 383--425 ISSN 0034-6861
  \urlprefix\url{http://link.aps.org/doi/10.1103/RevModPhys.82.383}

\bibitem{Limandri2014a}
Limandri S, {De Vera} P, Fadanelli R~C, Nagamine L~C, Mello A, Garcia-Molina R,
  Behar M and Abril I 2014 {\em Physical Review E - Statistical, Nonlinear, and
  Soft Matter Physics\/} {\bf 89} 022703

\bibitem{deVera2018}
de~Vera P, Abril I and Garcia-Molina R 2018 {\em Radiation Research\/} {\bf
  190} 282--297

\bibitem{deVera2014b}
{de Vera} P, Abril I and Garcia-Molina R 2014 {\em Applied Radiation and
  Isotopes\/} {\bf 83} 122--127 ISSN 09698043

\bibitem{Bagheri2019}
Bagheri R, Moghaddam A~K, Azadbakht B, Akbari M~R and Shirmardi S~P 2019 {\em
  Nuclear Science and Techniques\/} {\bf 30}(2) 31

\bibitem{Burin2023}
Burin A, Branco I and Yoriyaz H 2023 {\em Radiation Physics and Chemistry\/}
  {\bf 203} 110606

\bibitem{ICRU46}
ICRU 1992 {\em {Report 46 - Photon, electron, proton and neutron interaction
  data for body tissues}\/} (Bethesda, Maryland: International Commission on
  Radiation Units and Meassurements)

\bibitem{ICRU1998}
ICRU 1998 {\em {Report 59 - Clinical Proton Dosimetry. Part I: Beam Production,
  Beam Delivery and Measurement of Absorbed Dose}\/} (Bethesda, Maryland:
  International Commission on Radiation Units and Measurements)

\bibitem{Shimizu2009}
Shimizu M, Kaneda M, Hayakawa T, Tsuchida H and Itoh A 2009 {\em Nuclear
  Instruments and Methods in Physics Research, Section B: Beam Interactions
  with Materials and Atoms\/} {\bf 267} 2667--2670

\bibitem{Shimizu2010}
Shimizu M, Hayakawa T, Kaneda M, Tsuchida H and Itoh A 2010 {\em Vacuum\/} {\bf
  84} 1002--1004

\bibitem{Siiskonen2011}
Siiskonen T, Kettunen H, Per{\"{a}}j{\"{a}}rvi K, Javanainen A, Rossi M,
  Trzaska W~H, Turunen J and Virtanen A 2011 {\em Physics in Medicine and
  Biology\/} {\bf 56} 2367--2374

\bibitem{Dingfelder2000}
Dingfelder M, Inokuti M and Paretzke H~G 2000 {\em Radiation Physics and
  Chemistry\/} {\bf 59} 255--275 ISSN 0969806X
  \urlprefix\url{http://linkinghub.elsevier.com/retrieve/pii/S0969806X00002632}

\bibitem{Akkerman2001}
Akkerman A 2001 {\em Radiation Physics and Chemistry\/} {\bf 61} 333--335 ISSN
  0969806X

\bibitem{Date2006}
Date H, Sutherland K~L, Hayashi T, Matsuzaki Y and Kiyanagi Y 2006 {\em
  Radiation Physics and Chemistry\/} {\bf 75} 179--187 ISSN 0969806X
  \urlprefix\url{http://linkinghub.elsevier.com/retrieve/pii/S0969806X05002598}

\bibitem{Emfietzoglou2009}
Emfietzoglou D, Garcia-Molina R, Kyriakou I, Abril I and Nikjoo H 2009 {\em
  Physics in Medicine and Biology\/} {\bf 54} 3451 ISSN 0031-9155

\bibitem{GarciaMolina2009}
Garcia-Molina R, Abril I, Denton C~D, Heredia-Avalos S, Kyriakou I and
  Emfietzoglou D 2009 {\em Nuclear Instruments and Methods in Physics Research
  B\/} {\bf 267} 2647--2652

\bibitem{Tan2004}
Tan Z, Xia Y, Zhao M, Liu X, Li F, Huang B and Ji Y 2004 {\em Nuclear
  Instruments and Methods in Physics Research Section B: Beam Interactions with
  Materials and Atoms\/} {\bf 222} 27--43

\bibitem{Koehler1965}
Koehler A~M, Dickinson J~G and Preston W~M 1965 {\em Radiation Research\/} {\bf
  26} 334--342

\bibitem{ICRU49}
ICRU 1993 {\em {Report 49 - Stopping powers and ranges for protons and alpha
  particles}\/} (Bethesda, Maryland: International Commission on Radiation
  Units and Measurements)

\bibitem{Berger2005}
Berger M~J, Coursey J~S, Zucker M~A and Chang J 2005 {ESTAR, PSTAR, and ASTAR:
  Computer programs for calculating stopping power and range tables for
  electrons, protons, and helium ions}
  \urlprefix\url{http://physics.nist.gov/Star}

\bibitem{GarciaMolina2011}
Garcia-Molina R, Abril I, Heredia-Avalos S, Kyriakou I and Emfietzoglou D 2011
  {\em Physics in Medicine and Biology\/} {\bf 56} 6475--6493 ISSN 00319155

\bibitem{GarciaMolina2012SpringerScience}
Garcia-Molina R, Abril I, Kyriakou I and Emfietzoglou D 2012 {Energy Loss of
  Swift Protons in Liquid Water: Role of Optical Data Input and Extension
  Algorithms} {\em Radiation Damage in Biomolecular Systems\/} ed {Garc{\'{i}}a
  G{\'{o}}mez-Tejedor} G and Fuss M~C (Dordrecht: Springer Science+Business
  Media B.V.) chap~15 ISBN 978-94-007-2563-8

\bibitem{Abril1998}
Abril I, Garcia-Molina R, Denton C~D, P{\'{e}}rez-P{\'{e}}rez F~J and Arista
  N~R 1998 {\em Physical Review A - Atomic, Molecular, and Optical Physics\/}
  {\bf 58} 357--366

\bibitem{HerediaAvalos2005a}
Heredia-Avalos S, Garcia-Molina R, Fern{\'{a}}ndez-Varea J~M and Abril I 2005
  {\em Physical Review A - Atomic, Molecular, and Optical Physics\/} {\bf 72}
  052902 ISSN 10502947

\bibitem{vasiliev2010tissue}
Vasiliev V~N, Kostjuchenko V~I, Riazantsev O~B and Khaybullin V~G 2010 {\em
  arXiv\/} {\bf 1005.4389} [physics.med--ph]

\bibitem{Fermi1940}
Fermi E 1940 {\em Physical Review\/} {\bf 57} 485--493

\bibitem{Lindhard1954}
Lindhard J 1954 {\em Det Kongelige Danske Videnskabernes Selskab.
  Matematisk-fysiske Meddelelser\/} {\bf 28}(8) 1--57

\bibitem{Ritchie1957}
Ritchie R~H 1957 {\em Physical Review\/} {\bf 106} 874--881

\bibitem{Emfietzoglou2012}
Emfietzoglou D, Kyriakou I, Abril I, Garcia-Molina R and Nikjoo H 2012 {\em
  International Journal of Radiation Biology\/} {\bf 88} 22--28 ISSN 09553002

\bibitem{Mermin1970}
Mermin N~D 1970 {\em Physical Review B - Condensed Matter and Materials
  Physics\/} {\bf 1} 2362--2363

\bibitem{Abril2010a}
Abril I, Denton C~D, de~Vera P, Kyriakou I, Emfietzoglou D and Garcia-Molina R
  2010 {\em Nuclear Instruments and Methods in Physics Research Section B: Beam
  Interactions with Materials and Atoms\/} {\bf 268} 1763--1767 ISSN 0168583X
  \urlprefix\url{http://linkinghub.elsevier.com/retrieve/pii/S0168583X10001643}

\bibitem{Watanabe1997}
Watanabe N, Hayashi H and Udagawa Y 1997 {\em Bulletin of the Chemical Society
  of Japan\/} {\bf 70} 719--726

\bibitem{Hayashi2000}
Hayashi H, Watanabe N, Udagawa Y and Kao C~C 2000 {\em Proceedings of the
  National Academy of Sciences of the United States of America\/} {\bf 97}
  6264--6266 ISSN 00278424

\bibitem{Emfietzoglou2008}
Emfietzoglou D, Abril I, Garcia-Molina R, Petsalakis I~D, Nikjoo H, Kyriakou I
  and Pathak A 2008 {\em Nuclear Instruments and Methods in Physics Research,
  Section B\/} {\bf 266} 1154--1161 ISSN 0168583X

\bibitem{Rulis2004}
Rulis P, Ouyang L and Ching W 2004 {\em Physical Review B - Condensed Matter
  and Materials Physics\/} {\bf 70} 0155104

\bibitem{deVera2013PRL}
{de Vera} P, Garcia-Molina R, Abril I and Solov'yov A~V 2013 {\em Physical
  Review Letters\/} {\bf 110} 148104 ISSN 00319007

\bibitem{deVera2015}
{de}~Vera P, Garcia-Molina R and Abril I 2015 {\em Physical Review Letters\/}
  {\bf 114} 018101 ISSN 0031-9007
  \urlprefix\url{http://link.aps.org/doi/10.1103/PhysRevLett.114.018101}

\bibitem{Sigmund2009}
Sigmund P, Schinner A and Paul H 2009 {\em Journal of the ICRU\/} {\bf 5} 1--10

\bibitem{GarciaMolina2012WorldScientific}
Garcia-Molina R, Abril I, {de Vera} P, Kyriakou I and Emfietzoglou D 2012
  {Proton Beam Irradiation of Liquid Water: A Combined Molecular Dynamics and
  Monte Carlo Simulation Study of the Bragg Peak Profile} {\em Fast Ion-Atom
  and Ion-Molecule Collisions\/} ed Belki\'c D (Singapore: World Scientific
  Publishing Company) chap~8, pp 271--304

\bibitem{GarciaMolina2014}
Garcia-Molina R, Abril I, de~Vera P, Kyriakou I and Emfietzoglou D 2014 {\em
  Applied Radiation and Isotopes\/} {\bf 83} 109--114 ISSN 09698043
  \urlprefix\url{http://dx.doi.org/10.1016/j.apradiso.2013.01.006}

\bibitem{Allen1989}
Allen M~P and Tildesley D~J 1989 {\em {Computer simulation of liquids}\/}
  (Oxford: Oxford University Press)

\bibitem{Box1958}
Box G~E~P and Muller M~E 1958 {\em Ann. Math. Stat.\/} {\bf 29} 610--611

\bibitem{ICRU63}
ICRU 2000 {\em {Report 63 - Nuclear Data for Neutron and Proton Radiotherapy
  and for Radiation Protection}\/} (Bethesda, Maryland: International
  Commission on Radiation Units and Meassurements)

\bibitem{Medin1997}
Medin J and Andreo P 1997 {\em Physics in Medicine and Biology\/} {\bf 42}
  89--105

\bibitem{Schiwietz2001}
Schiwietz G and Grande P~L 2001 {\em Nuclear Instruments and Methods in Physics
  Research Section B: Beam Interactions with Materials and Atoms\/} {\bf
  175-177} 125--131

\bibitem{Wenzel1952}
Wenzel W~A and Whaling W 1952 {\em Physical Review\/} {\bf 87} 499--503

\bibitem{Andrews1977}
Andrews D~A and Newton G 1977 {\em Journal of Physics D\/} {\bf 10} 845--850

\bibitem{Bauer1994}
Bauer P, Kaferbock W and Necas V 1994 {\em Nuclear Instruments and Methods in
  Physics Research Section B: Beam Interactions with Materials and Atoms\/}
  {\bf 93} 132--136

\bibitem{SRIM2013}
Ziegler J~F 2013 {SRIM - The Stopping and Range of Ions in Matter}
  \url{http://www.srim.org}

\bibitem{GarciaMolina2013NIMB}
Garcia-Molina R, Abril I, {de Vera} P and Paul H 2013 {\em Nuclear Instruments
  and Methods in Physics Research, Section B\/} {\bf 299} 51--53 ISSN 0168583X
  \urlprefix\url{http://dx.doi.org/10.1016/j.nimb.2013.01.038}

\bibitem{Sigmund2006}
Sigmund P 2006 {\em Particle Penetration and Radiation Effects. General Aspects
  and Stopping of Swift Point Charges\/} (Berlin Heidelberg: Springer-Verlag)
  ISBN 978-3-540-31713-5

\bibitem{Zhang2011}
Zhang X, Liu W, Li Y, Li X, Quan M, Mohan R, Anand A, Sahoo N, Gillin M and Zhu
  X~R 2011 {\em Physics in Medicine and Biology\/} {\bf 56} 7725--7735

\end{thebibliography}
\bibliographystyle{iopart-num} 


\clearpage





\end{document}